\def\year{2015}
\pdfoutput=1
\documentclass[letterpaper]{article}
\usepackage{aaai}
\usepackage{times}
\usepackage{helvet}
\usepackage{courier}
\usepackage{graphicx}
\begin{document}
%
\title{The Residence History Inference Problem}

\author{
\begin{tabular}{cc}
Derek Ruths & Caitrin Armstrong\\
\small{derek.ruths@mcgill.ca} & \small{caitrin.armstrong@mail.mcgill.ca}
\end{tabular}\\
School of Computer Science\\
McGill University
}

\maketitle
\begin{abstract}
\begin{quote}
The use of online user traces for studies of human mobility has received significant attention in recent years.  This growing body of work, and the more general importance of human migration patterns to government and industry, motivates the need for a formalized approach to the computational modeling of human mobility - in particular how and when individuals change their place of residence - from online traces. Prior work on this topic has skirted the underlying computational modeling of residence inference, focusing on migration patterns themselves.  As a result, to our knowledge, all prior work has employed heuristics to compute something like residence histories.  Here, we formalize the {\it residence assignment problem}, which seeks, under constraints associated with the minimum length-of-stay at a residence, the most parsimonious sequence of residence periods and places that explains the movement history of an individual.  Here we provide an exact solution for this problem and establish its algorithmic complexity.  Because the calculation of optimal residence histories (under the assumptions of the model) is tractable, we believe that this method will be a valuable tool for future work on this topic.
\end{quote}
\end{abstract}

\section{Introduction}
Whether for short-term travel or longer-term migration, the movements of human populations impact culture, language and economics in fundamental and lasting ways.  As a result, human migration patterns are a topic of intense interest to scholars, governments, human rights groups, and other organizations. 

Migration has always been challenging to study due to a lack of high resolution and up-to-date data.  What data does exist, in the form of census statistics or survey results, are static, have a bias towards legally-protected populations, lack valuable descriptive population demographics, and lag the current population by months, often years. New solutions are needed.

The widespread adoption of online social platforms such as Twitter, Facebook, WhatsApp, and Skype present exciting new ways of measuring and characterizing migration patterns.  Such platforms obtain information about a user's historical locations through two major means.  A user may self-report their location (e.g., ``Having fun with @jsmith in NYC!'') or even moves (e.g., ``Just arrived in London. \#sotired'').  A user's location is also implied by the IP address their computer or phone is using --- public IP addresses are easily resolvable to latitude-longitude coordinates that, while rarely exact, can bound the user's location by several miles.  Over time, either source of data can provide a history of the locations the user has stayed at.

A number of existing research efforts have used such digitally-constructed location histories to infer and then study country- or region-level {\it residence histories}: the dates at which a user changed their home and where they moved to \cite{zagheni2014inferring,state2014migration}. Intuitively, residence may seem like a straightforward concept. One simply ``lives'' wherever they spend the night and mobility events translate neatly into residence changes. And yet we do not consider a short trip as changing where the traveler lives. Thus, even though location and residence are deeply intertwined notions, they are not the same. Indeed, in ``the lay meaning of residence, [it] certainly arises after one has lived in a place for a reasonably long time; it may also arise after a comparatively short stay, or even immediately upon arrival, provided that one intends to remain there for a considerable period in the future.'' \cite{reese1952elusive}.

Thus, the {\it residence history inference problem} that these existing studies have had to solve is, by its nature, ill-defined.  As we will see, once properly defined, the problem is also a non-trivial one.  To our knowledge, every prior study of residence has employed a heuristic in order to infer residence histories.  Heuristics are naturally employed when the exact solution is intractable or impossible to obtain. So we might take the exclusive use of heuristics to be an indication that, at some point, the residence history inference problem has been proven NP-complete or worse.  Surprisingly, no such study has been done: the formal properties of the residence history inference problem are entirely unknown.  The practical implication of this is that an entire field of research may be resorting to heuristics, satisfying themselves with approximate residence histories, when the exact histories are, in fact, perfectly obtainable.  This is the subject of the present work.

In this paper we focus on specifying and analyzing this \textit{residence history inference problem}: determining an individual's historical residence locations and time-intervals from a list of time-stamped locations at which that individual has been. In this work, we make two key contributions. 

First we formalize the problem and show that it is, in fact, quite tractable to solve.

Second, we provide an exact, polynomial-time algorithm solution.

Given continued and growing interest in migration studies, particularly using online and cell phone traces, the formalization of this problem and the exact solution we provide will put future quantitative work on migration and mobility on stronger, more theoretically sound footing.

In order to make our findings useful, we have also released a software tool that implements our algorithm\footnote{\url{https://github.com/networkdynamics/resin}}.

In the remaining sections we first review past approaches to quantifying migration from trace data.  We then motivate and provide a formal definition of the residence inference problem. Finally, we offer an exact solution to the problem and prove its correctness and its efficiency.

\section{Prior Work}
Human mobility includes a range of phenomena ranging from daily commuting routines to urbanization trends that span decades.  In this paper, we are focused on migration - a particular kind of mobility that captures when individuals change their durable home or base of operation.

During the past two decades, a substantial and growing body of research has taken a scientific lens to human migration patterns.  This work has been fueled, in part, by the emergence of very large human trace datasets created by new technologies such as the Internet and cellphone networks.  These datasets uniquely capture the distinct activities of individuals (as opposed to groups or communities). Scientists have long recognized the promise these datasets hold for the advancement of our understanding of basic human social processes \cite{lazer2009computational}. 

In particular, computational social science has recently begun showing promise for the advancement of research into human migration patterns using large human trace datasets. Cellphone call-record data has been particularly helpful in studying dynamics up to the national level. In a study of an entire country's call records, \cite{phithakkitnukoon2012socio} examined the relationship between patterns of internal migration in Portugal and the evolution of social networks. Similarly, \cite{blumenstock2012inferring} used call-record data to create estimates of internal migrations for Rwanda. \cite{zagheni2012you} were the first to show that IP geolocation can be used to create country-dyad-level estimates of migration. \cite{weber2013studying} advanced this method further by producing full country-to-country migration and tourism matrices from IP geolocation data. \cite{zagheni2014inferring} showed how Twitter data could be used to generate estimates of both international and internal migration patterns. \cite{state2014migration} used data from the professional network LinkedIn to produce estimates of highly-skilled migrant stocks across the world. 

In order to conduct such analysis, every one of these studies (and all studies like them) must identify migration events in individual activity records.  Since a migration event is, in effect, a change in the individual's ``home'', detection migration requires identifying the individual's residence at each point of time in the past.  Thus, in essence, every migration study must solve the residence history inference problem.  Existing work has, without exception, used heuristics that appeal to common-sense or legalistic notions of migration, thereby avoiding the task of formalizing the inference problem.

One such heuristic that has been used widely is the {\it modal location} approach (e.g., \cite{Fiorio2017}). This method consists of simply dividing the residence history into intervals of fixed length, and assigning the modal location during each interval as the residence for that interval. This approach has the advantage of low computational complexity, offering a linear-time solution. Nonetheless, there are conceivable situations for which this approach would disagree with the exact solution: for instance, a user could spend 16 days in location A, move to location B and spend 14 days there, return to location A for 16 days, and then spend 44 days in location B. Using 30-day intervals would assign the user's residence to location A for the first two periods, and to location B only for the last period. A move to location B would thus be detected only two months after the user actually changed residence. Despite these shortcomings, it's worth noting that even in this contrived example the heuristic catches up with the user's real location history, as the correct residence is eventually assigned.

\section{The Residence History Inference Problem}
In this section, we formalize, for the first time, the problem which prior work has implicitly approached using heuristic methods.

The residence assignment problem seeks the most likely set of residence locations and intervals (hereafter, the {\it residence history}) that explains a series of time-stamped observations of a user at different locations, which we call an individual's {\it location history}.

To formalize this problem, we begin by breaking the time period of interest into time unit intervals within which we can assign the person to one specific location.  Here, for the purpose of clarity and concreteness, we will assume intervals to be days, but, the temporal scale is a parameter of the model which does not impact the algorithmic properties.

Our observation data (hereafter, {\it location history}), $H = \langle h_1, h_2, ..., h_n \rangle$, provides a location for each day in our time period.  So, for example, if our time period of interest is a year, then $|H| = 365$.

We seek to infer from this location history the locations and intervals during which the user {\it resided} at different places.  We represent this, like the observational data, as a sequence of locations, $R = \langle r_1, r_2, ..., r_n \rangle$, with one location per day.  Notice that when $r_i \not= r_{i+1}$, the user has moved residences on day $i+1$. Similarly, when $r_i \not= h_i$, the user is traveling away from home.

This problem is effectively a latent attribute inference task where the observational data is giving signal about where the user lives.  Thus, we are interested in the residence histories that do the best job of explaining the locations observed in $H$.

We submit that a strong location-based signature of residence is time spent in that location: a person who intends to live in a place will eventually end up spending significant time there.  This is the intuition that informs tax and immigration law as well as numerous studies of migration --- and we employ it here.



In this way of thinking, the best residence history will be the one in which the individual spends the most days at their residence locations.  
Generalizing this idea, the best residence history will minimize 
\begin{equation} \label{eqn:more}
	\sum_{i = 1}^{n} \gamma(h_i,r_i)
\end{equation}

where $\gamma(x,y) = 1 \mbox{ iff } x \not= y$.

Selecting a residence history based on this single criterion admits a trivial solution: always assert that the user resides wherever we observe them ($\forall i, h_i = r_i$). Residence histories are typically more complex as people take trips which do not correspond to changes in residence.

We need to introduce a second criterion which penalizes solutions that create too many residence changes --- effectively over-fitting the location history.  Myriad approaches to modeling and legislating residence suggest the use of a minimum residence interval length (e.g., 90 days for the UN, 183 days for international tax law).  This acknowledges the practical reality that, while one may intend to reside in a place, considerable time spent in that location, to the exclusion of other places, constitutes evidence of that initial (and continuing) intention.  Thus, in addition to the inequality in Equation \ref{eqn:more}, we also require that each residence period last at least $\rho$ contiguous days.

\begin{figure}[t]

\fbox{
\begin{minipage}{3in}
\begin{center}
{\it {\large The Residence Inference Problem}}\\
\end{center}
\noindent {\bf Inputs:}
\begin{itemize}
	\item Location history, $H = \langle h_1, h_2, ..., h_n \rangle$
	\item Minimum residence interval length, $\rho$
\end{itemize}
\vspace{0.1in}
\noindent {\bf Output:} Residence history $R = \langle r_1, r_2, ..., r_n \rangle$ such that
\begin{itemize}
	\item for all $i>1$ where $r_{i+1} \not= r_i$, \\$\forall 0 \leq j < \rho$, $r_{i-j} = r_i$, and
	\item for any other history $R'$,\\ $\sum_{i = 1}^{n} \gamma(h_i,r_i) \leq \sum_{i = 1}^{n} \gamma(h_i,r'_i)$
\end{itemize}

\end{minipage}
}

\caption{The residence assignment problem. \label{fig:def}}
\end{figure}

This yields the problem definition given in Figure \ref{fig:def}.

Note that other alternatives to the minimum residence interval might be taken.  For example, we could directly penalize longer residence intervals by using the objective function $\min m\hat{D}$, where $m$ is the number of residence intervals and $\hat{D}$ is the total number of days the individual spent away from their residence locations.  A likelihood-based approach might also be used with a objective function having the form $\max P(d_{i+1}-d_i)D$, where $P(i)$ is the probability of a residence period with interval $i$ and $D$ is the number of days the individual spent at their residence locations.

Both of these alternative formulations present the serious challenge of learning weighting and other parameters from labeled data.  As labeled migration data is very scarce, here we focus on the original formalization (which uses the minimum residence interval length, $\rho$), and identity these other criteria as promising directions for future work.

\section{An Exact Solution}
A generalized solution to the residence inference problem proceeds as follows: every observed location change, $h_i \not= h_{i+1}$, is a possible change in residence (i.e., a possible $r_i \not= r_{i+1}$ in the final solution).  Thus, if we have a residence history $R_i$ constructed up to day $i$, then the change in location requires us to consider two derivative residence histories $R_{i,1}$ and $R_{i,2}$: in the first, the location change was {\it not} a residence change; in the second, it {\it was} a residence change.  Every $i$ at which $h_i \not= h_{i+1}$ induces such a branching on all existing solutions up to $i$.  Once the final time interval has been processed, the total days-away-from-residence is computed for each candidate solution and the residence history with the lowest score is returned.  



Crucially, this approach yields a branching exploration of the solution space 
which yields, in the worst case, a number of solutions that is exponential in $k$, the number of observed moves (i.e., $k = |\{i : h_i \not= h_{i+1}\}|$).  As $k$ will be somewhat correlated to $|L|$, this suggests that this approach becomes computationally intractable with longer observation periods.  Naturally, we must see if we can do better.

Happily, the additive cost involved in the objective function (see Equation \ref{eqn:more}), admits a computationally tractable dynamic programming approach.  This is because an observation on day $i$ cannot affect the solution up through day $i-\rho$. One way to think about this is that the minimum residence period defines the number of subsequent days that inform the residence location at time $i$.

Because the addition of another observation at time $i$ does not change the optimal solution to the sub-problem to days $1$ through $i-\rho$, we can formulate this using the following dynamic programming function:

\begin{equation}
	A(i,l) = \min_{l' \in (L-\{l\}),t \leq i-\rho} \big( A(t,l') + \sum_{j = t+1}^i \gamma(l,h_j) \big).
\end{equation}

Here $A(i,p)$ is the minimum number of days the user must have been away from her residence for the time interval $[1,i]$ with a residence history ending with location $p$.  The optimal residence history can be constructed from the subproblems embedded in the final solution: the sequence of $A(i,p)$'s in the solution indicate the time at which and destination to which the user moved.

The time complexity of this approach is $O(|H|^2|L|^2)$ where $|H|$ is the number of intervals in the location history and $L = \mbox{Set}(H)$ is the set of locations that appear in the history.

\paragraph{Optimizing by time-warping the location history.} Notice that there is never a reason to infer a change in residence when the user has not changed location.  Such a change would either (1) be to the current location, $h_i$, in which case it would have been better to have changed residence when the user first arrived at that location, or (2) be to a different location, in which case a change of residence at this time does not decrease the number of away days being accumulated.  As a result, one simple optimization is to construct time intervals so that they are of variable duration and end when the user is observed in a different location.  This would give the {\it time-warped history}, $W = \langle (w_1,c_1), (w_2,c_2), ..., (w_k,c_k) \rangle$, where $w_i$ is the location of the user and $c_i$ is the number of days the user is in that location.  We can revise the dynamic program to:

\begin{equation}
	A(i,l) = \min_{l' \in (L-\{l\}),t \in Q(i)} \big( A(t,l') + \sum_{j = t+1}^i \gamma(l,w_j)c_j \big),
\end{equation}

where $Q(i) = \{ j : 0<j<i \mbox{ and } \sum_{k=j}^i c_k \geq \rho \}$ is the set of indices into the time-warped history that start at least $\rho$ days before time interval $i$. The key difference here is that we are no longer looping over all time intervals (e.g., days), but rather over the time-warped intervals, skipping over periods when the user did not change location.

This approach improves the time complexity of the algorithm to $O(|W|^2|L|^2)$. Since $|W| \leq |H|$, this avoids a necessary increase in computational cost with longer location histories (only an increase in the number of moves will require additional effort).

\paragraph{Different cost functions.} Notice that this dynamic programming solution is possible because of the independence of earlier subsolutions from later observations.  In particular, the dynamic program separates previous residences from the addition of a new residence (location $l$).  This same approach will work for other cost functions (e.g., besides $\gamma(x,y)$) that update the score of the solution based only on the duration/attributes of the current residence interval.  For example, a likelihood function $P(d)$ which penalizes the residence history based on its length would only involve replacing the summation in the dynamic program, which does not affect the complexity of the problem.

\section{Conclusion}
In this paper, we have made two important contributions to the active field of migration studies using online trace data.  First, we have formalized the {\it residence history inference problem}, the core computational task involved in deriving migration events from social trace data.  Up until now, prior work has employed heuristics which have informally engaged with this problem without clearly stating the singular problem they all have been seeking to solve.  Our second contribution is an efficient algorithm that will always infer the optimal residence history from a user's online location trace.

Our hope is that this exact solution will provide researchers an effective tool for conducting future studies of migration using online location trace data.  Furthermore, as we have pointed out, a number of alternative formulations of the residence history inference problem exist which are likely both more complex and more expressive.  Our hope is that these and related formulations will yield promising ground for further methodological work on this important topic.

\bibliographystyle{aaai}
\bibliography{biblio}

\end{document}